\documentclass[doublecol]{epl2}
\usepackage{graphicx}

\title{
Pattern Formation and Transport for Externally Driven Active Matter 
on Periodic Substrates
} 
\author{
C. Reichhardt and C. J. O. Reichhardt 
} 
\institute{
Theoretical Division and Center for Nonlinear Studies,
Los Alamos National Laboratory, Los Alamos, New Mexico 87545, USA\\ 
} 

\abstract{
We investigate the transport of interacting active run-and-tumble particles
moving under an external drift force through a
periodic array of obstacles for increasing
drive amplitudes. 
For high activity where the system forms  
a motility induced phase separated state, there
are several distinct dynamic phases including a low drive 
pinned cluster phase, an intermediate uniform fluid, and a higher drive stripe
crystal state. The transitions between the phases
are correlated with signatures in the transport curves,
differential mobility, and power spectra of the velocity
fluctuations. In contrast, in the low activity regime the transport
curves and power spectra
undergo little change as a function of drive. We argue that in the 
high activity limit,
the behavior is similar to that of driven solids on periodic substrates,
while in the low activity limit the system behaves like a driven fluid. 
}
\begin{document}
\maketitle

\section{Introduction}
There is a wide variety of systems that can be described
as an assembly
of interacting particles driven over a periodic substrate or arrays of
obstacles \cite{Reichhardt17}. Examples of this type of system  include
vortices in type-II superconductors \cite{Baert95,Harada96,Martin97,Reichhardt98},
skyrmions \cite{Reichhardt22}, 
colloids \cite{Reichhardt02a,Brunner02},
Bose-Einstein condensates \cite{Tung06}, liquid crystals \cite{Duzgun20},
and frictional systems \cite{Vanossi13}.
When solids or crystals
are driven over a periodic substrate,
distinct dynamical phases appear such as a low drive pinned state,
a soliton state consisting of pinned and moving
particles, plastically deforming states, fluidized states,
and moving smectics 
\cite{Reichhardt97,Gutierrez09,Bohlein12,Vanossi12,McDermott13a,Hasnain14,Reichhardt17}.
Transitions between different moving states can be associated
with distinct types of fluctuations 
and changes in velocity,
leading to jumps or even dips in the velocity-force 
and differential mobility curves.
In contrast, for fluids driven over periodic substrates,
fewer features appear
and the response is often linear with drive \cite{Reichhardt17}.

Another type of collective particle system that is currently 
being studied is active matter or self-driven particles
\cite{Marchetti13,Bechinger16,Gompper21}. In the absence of 
a substrate, such systems show interesting phenomena
such as motility induced
phase separation (MIPS),
where the particles can self-cluster even when all of the
particle-particle interactions are repulsive
\cite{Fily12,Redner13,Palacci13, Buttinoni13,Reichhardt15,Cates15}.
Once a substrate or wall is added, 
a variety of behaviors
can appear that are not found in passive particle systems,
including novel pressure effects \cite{Nikola16},
ratchet effects \cite{DiLeonardo10,Reichhardt13,Ai14,Granek22},
and nucleation or wetting phases at walls or obstacles
\cite{Sandor17b,Sepulveda17,Neta21}. 

Active matter systems coupled to  periodic  substrates have been studied 
with a variety of simulations
and experiments in both the single particle and collective limits 
\cite{Volpe11,Reichhardt20,BrunCosmeBruny20,Reichhardt22a,Nabil22,Chopra22}. 
Computational work on interacting systems
has even shown that an active matter
commensuration or Mott effect can occur
in which for some fillings
the active particles form an ordered MIPS
state that is strongly pinned to the substrate, while for other fillings
there is
a more mobile active cluster glass phase \cite{Reichhardt21}.  
In systems with random or periodic substrates,
increasing the activity may initially improve the ability of
active particles to move in response to an applied drift force,
but once the activity
becomes high enough to induce
the formation of clustering phases,
a clogging of the flow occurs
\cite{Reichhardt14,Reichhardt21a}. 
Most studies of interacting
active matter on periodic substrates have been
performed in the limit where 
there is either no drift force or where the external drift  force
is small;
however, it is known from
studies of nonactive systems
such as driven solids
on periodic substrates \cite{Reichhardt17} 
that a variety of other flow phases are possible at higher drives. 

In this work we consider collectively interacting
active particles driven over a periodic substrate in both
the high activity regime where the system forms
a MIPS
state and
in the low activity or Brownian fluid regime.
For high activity,
at low drives the system
is in an active clogged or pinned phase as previously
observed \cite{Reichhardt21a},
while with increasing drive
a depinning transition occurs into a uniform fluid, followed at
high drives by 
a dynamical reordering into a phase separated  stripe
crystal.
The onset of the different phases can be
detected via
peaks in the differential mobility as
well as non-monotonic behavior of the 
clustering.
We also show
that velocity noise spectra can be used
to characterize the
dynamics.
The velocity noise is large and of $1/f$ form
in the coexistence regime near depinning,
while the fluid phase and the stripe crystal
states have low noise power with white noise characteristics.
At low activities,
the velocity-force curves rapidly become
linear with increasing drive,
the noise power is low,
and there is a crossover
from a two-dimensional (2D) liquid to
a one-dimensional (1D) liquid at higher drives.
We argue that 
in the high activity limit, the system behaves
like a solid driven over a periodic
array, whereas for low activity
the system acts like a fluid with weak collective effects.

\section{Model}

We consider a 2D assembly of repulsive disks
with radius $R_{a}$ that interact with a square array
of obstacles, which are also modeled as repulsive disks with radius
$R_{\rm obs}$.
The system has periodic boundary conditions in the $x$ and $y$
directions, and
the equation of motion for disk $i$ is 
\begin{equation} 
        \eta \frac{d{\bf r}_i}{dt} = {\bf F}_{\rm inter}^i + 
	{\bf F}_{\rm sub}^i + {\bf F}_{m}^i + {\bf F}_{D} \ .
\end{equation}
Here the velocity of the disk
is ${\bf v}_{i} = {d {\bf r}_{i}}/{dt}$,
the location of the disk is ${\bf r}_{i}$,
and $\eta_d$  is the damping force,
which we set to $\eta_d=1.$
The particle-particle interaction forces
${\bf F}_{\rm inter}$ are given by
short range harmonic repulsion,
where we choose the interaction strength
such that the overlap of interacting disks
remains smaller than one percent of the disk radius. 
A similar
interaction force ${\bf F}_{\rm obs}$ is used for the obstacles,
which
are modeled as fixed position harmonic disks
placed in a square array with a lattice spacing of $a$. 
The motor force ${\bf F}_m$ represents run-and-tumble dynamics in which
each active disk experiences
a force of magnitude $F_{m}$ applied in a randomly chosen direction during
a run time of $\tau_{\rm run}$, after which
the disk randomly reorients and moves in a
different direction during the next running time interval.
We characterize the
system using the run length $l_r$,
defined to be the distance a disk would move during a single
run time in the absence of collisions
with other disks or obstacles,
$l_{r} = \tau_{\rm run}F_{m}$.
The overall density of the system is the area underneath both
the disks and the obstacles,
$\phi = N_{a}\pi R_{a}^2/L^2 + N_{obs}\pi R_{\rm obs}/L^2$,
where $L$ is the size of each side of the simulation box.
The driving force ${\bf F}_{D}=F_D{\bf \hat x}$ is applied in the
$x$-direction.
We measure the time series of the
velocity fluctuations in the direction
of drive,
$V_{x}=\sum_{i}^{N_a}{\bf v}_i \cdot {\bf \hat x}$,
as well as the time average of
this quantity, $\langle V_{x}\rangle$.
After applying 
a given drive $F_{D}$, we wait
a fixed number of time steps before
taking data in order to avoid any transient effects.
We measure $\langle V_x\rangle$ as a function of
varied $F_{D}$,
allowing us to construct velocity-force curves similar
to those studied 
for superconducting vortices, colloids,
and frictional systems \cite{Reichhardt17}.
In this work we fix the disk radius to $R_a=0.5$
and the obstacle
radius to $R_{\rm obs}=0.65$, and for most of the results
the obstacle lattice constant is set to $a=4.0$.

\section{Results} 
\begin{figure}
\onefigure[width=\columnwidth]{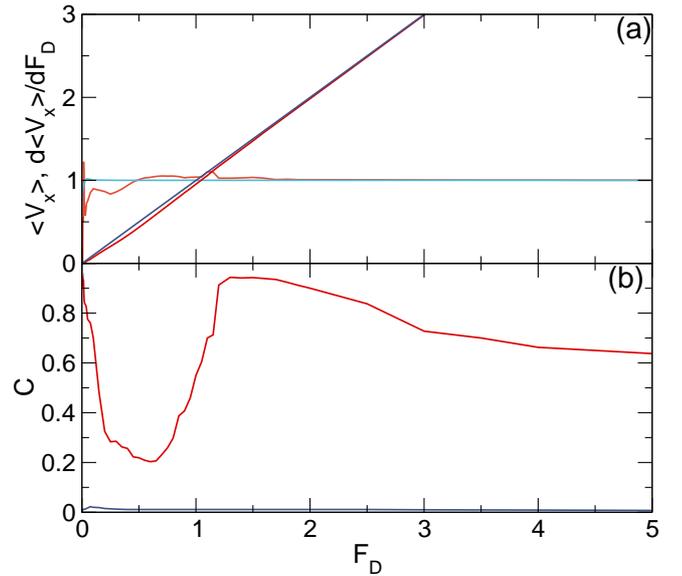}
\caption{(a) $V_{x}$ (dark red and dark blue curves) and
$d\langle V_{x}\rangle/dF_{D}$ (light red and light blue curves)
vs $F_{D}$ for run-and-tumble disks driven over a
periodic square obstacle array for running lengths of
$l_{r} = 0.02$ (dark and light blue)
and $l_{r}= 180$ (dark and light red).
Here $\phi = 0.5186$ and
the obstacle lattice spacing $a = 4.0$.
(b) The corresponding fraction
$C$ of particles in the largest cluster vs $F_D$.
}
\label{fig:1}
\end{figure}

We focus on the case where the total system density is
$\phi=0.5186$, well below
the jamming density  $\phi_J=0.9$ of a passive system \cite{Reichhardt14a}.
In the passive limit, there is
no clogging or jamming effect
and the velocity-force curves are linear for all values
of $F_{D}$, so any deviation from linearity that
we observe can be attributed to the activity.  
In Fig.~\ref{fig:1}(a) we plot $\langle V_{x}\rangle$ and
$d\langle V\rangle_{x}/dF_{D}$ versus $F_{D}$ for 
$l_{r} = 0.02$ in the low activity regime
and $l_{r}= 180$ in the high activity regime.  
Figure~\ref{fig:1}(b) shows the size of the largest cluster $C$ in the system,
obtained by detecting particles that are in direct contact with each other.
At $l_{r} = 0.02$,
the behavior is similar to that of Brownian particles and
the velocity-force curve has an almost completely linear dependence
on $F_D$, while
$C$ is always small. 
For the high activity system, $\langle V_{x}\rangle$
is depressed below the low activity curve
and the differential mobility $d\langle V_x\rangle/dF_D$ exhibits
several spikes and jumps.
The cluster measure $C$ is also nonmonotonic and
has a large value at low drives where the
system forms a clogged state in which most particles have aggregated into
a single cluster.
There is a drop in $C$ 
for $0.25 <F_{D} < 0.9$
when the system fluidizes,
a peak in $C$ near $F_{D} = 1.2$,
and a slow decrease in $C$ at higher drives when
the system reaches a stripe crystal state. 

\begin{figure}
\onefigure[width=\columnwidth]{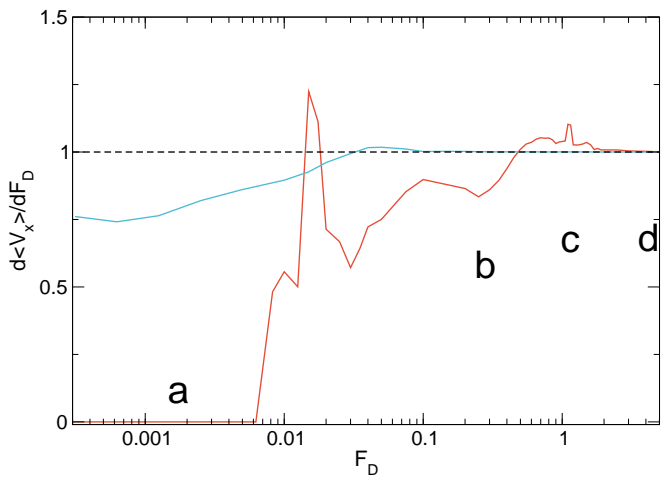}
\caption{
A plot on a linear-log scale of $d\langle V_x\rangle/dF_D$ vs $F_D$
from Fig.~\ref{fig:1}(a) at $\phi=0.5186$ and $a=4.0$
for $l_{r} = 0.02$ (light blue) and $l_{r} = 180$ (light red). 
The dashed line indicates the value of $d\langle V_x\rangle/dF_D$ that
would appear for a linear velocity-force curve.
The letters {\textbf a} through {\textbf d} indicate the values of $F_D$ at which the
images in Fig.~\ref{fig:3} were obtained for the $l_{r} = 180$ sample.
}
\label{fig:2}
\end{figure}

\begin{figure}
\onefigure[width=\columnwidth]{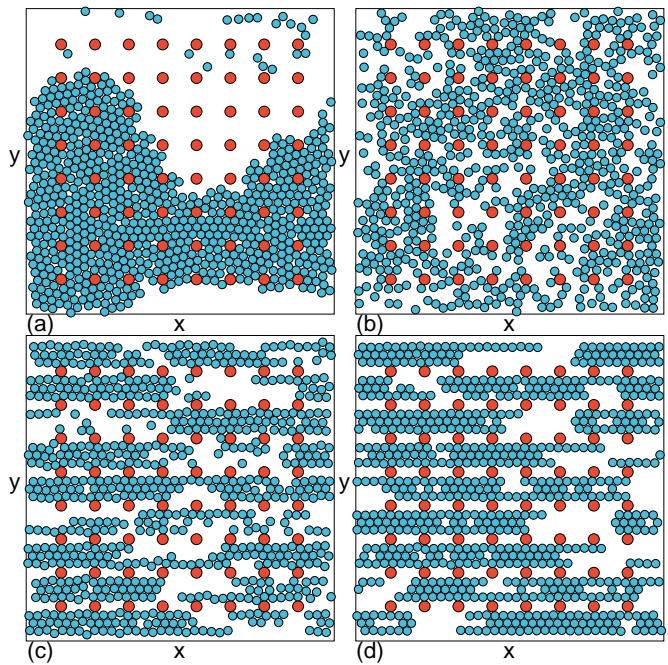}
\caption
{The particle positions (blue circles)
and obstacle locations (red circles) for the
$l_{r} = 180$ system
with $\phi=0.5186$ and $a=4.0$
at the drive values marked with letters
in Fig.~\ref{fig:2}.
(a) The clogged phase at $F_{D} = 0.0016$.
(b) The fluid phase at $F_{D} = 0.3$.
(c) The onset of density modulated stripe formation at $F_{D} = 1.0$.
(d) The density modulated stripe crystal at $F_{D} = 5.0$.
}
\label{fig:3}
\end{figure}

To more clearly display the nonlinearity in the transport for the two regimes, 
in Fig.~\ref{fig:2} we plot the same $d\langle V_x\rangle/dF_D$ versus $F_D$
curves from Fig.~\ref{fig:1} on
a linear-log scale.
The dashed line indicates the expected result
for a completely linear velocity-force curve, which is what we
observe for $l_{r} = 0$
when the particles are passively driven between the obstacles.
In the low activity regime,
$d\langle V_{x}\rangle/dF_{D}$ is less than one for $F_{D} < 0.05$.
Here 
the system forms a 2D fluid in which the particles move freely throughout
space. In this state, 
when a finite drive $F_{D}$ is applied, some of the particles
become temporarily trapped behind
the obstacles, leading to a reduction in $\langle V_x\rangle$ compared
to the purely passive system.
As $F_{D}$ increases, this trapping effect is reduced,
and for $F_{D} > 0.05$ the motion changes from 2D fluid flow to 1D channel
flow in which all of the particles remain confined between adjacent rows
of obstacles and the particle-obstacle interactions are minimized.
For the high activity regime,
there is a clogged phase for $F_{D} < 0.006$.
This state is illustrated for $F_D=0.0016$
in Fig.~\ref{fig:3}(a), 
where we highlight the mobile disk and obstacle positions. Here
the particles form a single large system spanning cluster where the
motion in the direction of drive is almost zero.
This activity-induced clogged phase is
similar to the one studied previously for active particles at a low drive
on periodic substrates \cite{Reichhardt21a,Reichhardt21}.
Near $F_{D} = 0.01$,
there is an effective depinning transition
in which the clogged state breaks apart
and the disks enter a moving fluid state, as shown
in Fig.~\ref{fig:3}(b) 
at $F_{D} = 0.3$.
Near $F_{D} = 1.0$, there is another peak in the
$d\langle V_x\rangle/dF_{D}$ curve corresponding
to a transition from the 2D fluid 
into a phase separated crystalline stripe phase,
illustrated  in Fig.3(c,d) for $F_{D} = 1.0$ and $5.0$, respectively.
At $F_{D} = 1.0$, there are some fluctuations in the structure and
the stripes intermittently break apart and reform,
while at $F_{D} = 5.0$, the stripes are stable. 
The stripes have an internal triangular crystalline ordering that is
oriented in the driving direction.

\begin{figure}
\onefigure[width=\columnwidth]{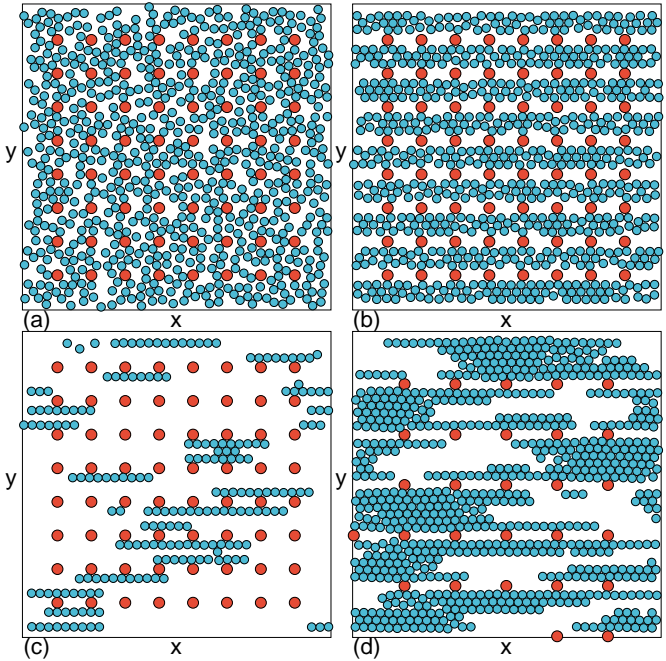}
\caption{
The particle positions (blue circles) and obstacle locations (red circles)
for the system from Fig.~\ref{fig:2} with $\phi=0.5186$ and $a=4.0$.
(a) $l_r=0.02$ and $F_D=0.00125$ in the 2D fluid phase.
(b) $l_r=0.02$ and $F_D=1.0$ in the 1D fluid phase.
(c) $l_{r} = 180$ at $F_{D} = 5.0$ in the stripe crystal
phase at $\phi = 0.2096$ where the number of active particles has
been reduced
and the system forms 1D chains. 
(d) $l_{r} = 180$ at $F_{D} = 5.0$ in the stripe crystal regime
at $\phi = 0.5186$ for a larger obstacle spacing
of $a = 6.0$, where the stripes are more 2D in nature and contain
up to five rows of particles each.
}
\label{fig:4}
\end{figure}

In Fig.~\ref{fig:4}(a) we show the particle locations for
the $l_r=0.02$ low activity system from Fig.~\ref{fig:2} 
at $F_{D} = 0.00125$ where the structure is a 2D fluid.
At $F_D=1.0$ in Fig.~\ref{fig:4}(b), 
the particles move in quasi-1D paths.
Unlike the high
activity case, the stripes remain liquidlike and there
is no local crystalline ordering or density
modulations in the direction of drive.

If the spacing $a$ between the obstacles is varied,
the depinning threshold can change and pass through maxima when
an integer number of particles are able to fit exactly
between the obstacles to form commensurate states.
When we vary the total system density $\phi$ by changing the density of
active particles, we find similar results in which there is a critical lowest
density below which the high activity system can no longer form a
clogged phase at low drives; however, the stripe cluster phase at higher drives
is robust. Additional features can occur, such as the
formation of stripes
with an integer number of rows, as shown in
Fig.~\ref{fig:4}(c) where the number of active particles 
in the system from Fig.~\ref{fig:3} has been
reduced to give a
total density of $\phi = 0.2096$. Here, 
the system forms a series of 1D chains at $F_{D} = 5.0$. 
For larger obstacle spacings $a$, the stripes become
more 2D in nature, as illustrated in Fig.~\ref{fig:4}(d)
for a system with the same density $\phi=0.5186$ as in Fig.~\ref{fig:3} but
with $a$ increased to $a=6.0$,
where up to five rows of particles can fit between
adjacent rows of obstacles. 

\begin{figure}
\onefigure[width=\columnwidth]{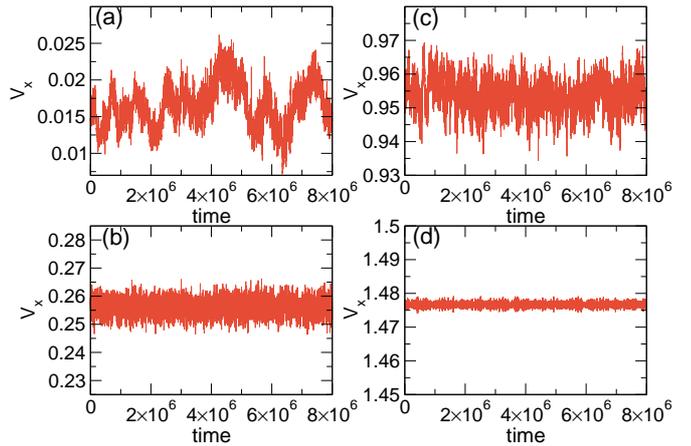}
\caption{
The time series of the $x$-direction velocities $V_x$ for the
system with $l_{r} = 180$, $\phi=0.5186$, and $a=4.0$.
(a) $F_{D} = 0.03$ just above the depinning transition from the
clogged phase.
(b) $F_{D} = 0.3$ in the fluid phase. 
(c) $F_{D} = 1.0$ where there is coexistence between the fluid
and a stripe crystal.
(d) $F_{D} = 1.5$ in the stripe crystal phase. 
}
\label{fig:5}
\end{figure}

Another method to characterize depinning and sliding dynamics
is by examining the velocity fluctuations and their power spectra.
For example,
during plastic depinning the
velocity fluctuations can be quite large since
large numbers of particles jump between pinned and moving states,
whereas in a moving fluid state, the particles
are continuously moving and the
velocity fluctuations are small \cite{Reichhardt17}.
In Fig.~\ref{fig:5} we 
plot time series 
of the $x$-direction velocities $V_x$ for the highly active system at 
$F_{D} = 0.03$, $0.3$, $1.0$, and $1.5$.
For $F_{D} = 0.03$ in Fig.~\ref{fig:5}(a), 
the system is just above the depinning
transition and there are strong velocity fluctuations.
In Fig.~\ref{fig:5}(b) at $F_{D} = 0.3$, 
the system is in a fluid phase and the
velocity fluctuations are less pronounced.
At $F_{D} = 1.0$ in Fig.~\ref{fig:5}(c),
the system is transitioning from the 2D fluid phase to a stripe
crystal and there is coexistence between the two states,
leading to enhanced velocity fluctuations,
while in the stripe crystal phase at $F_D=1.5$ in
Fig.~\ref{fig:5}(d), the velocity fluctuations
are strongly reduced.

\begin{figure}
\onefigure[width=\columnwidth]{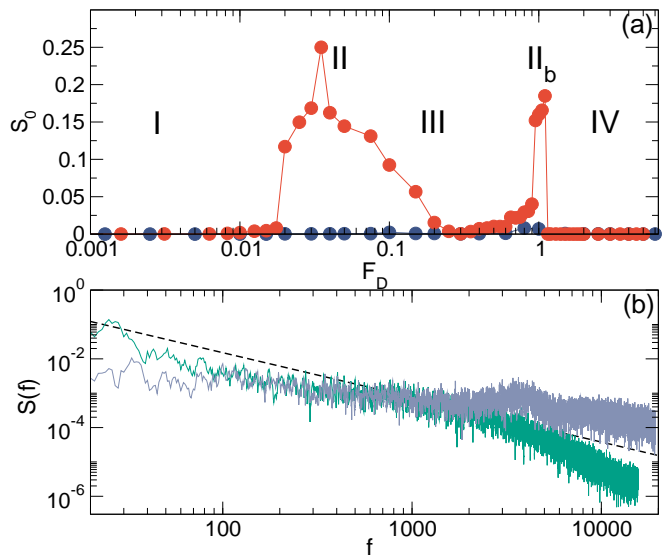}
\caption{
(a) The noise power $S_0$ over a fixed frequency range vs $F_{D}$ obtained
from the power spectra of the velocity time series for samples with
$\phi=0.5186$ and $a=4.0$ at a high activity of $l_r=180$ (red) and
a low activity of $l_r=0.02$ (blue).
For low activity, the noise power is always low.
The dynamic phases are:
I, clogged; II, coexistence between clogged and fluid;
III, fluid; II$_b$, coexistence between fluid and stripe crystal;
and IV, stripe crystal.
(b) The power spectra $S(f)$ vs frequency $f$
for the high activity $l_r=180$ sample in
phase II at $F_{D} = 0.03$ and in phase III
at $F_{D} = 0.3$. The dashed line is a power law fit 
with exponent $\alpha = -1.3$. 
}
\label{fig:6}
\end{figure}

To better characterize the changes in the
velocity fluctuations, we compute
the power spectrum of the velocity time series, 
$S(f) = |\int V_{x}(t)e^{-i2\pi f t}dt|$.
In Fig.~\ref{fig:6}(a) we plot the noise power $S_{0}$ versus $F_{D}$
at high and low activities.
We obtain $S_0$ by integrating $S(f)$ over a narrow window around
$f=10$.
The noise power is low for the low activity sample,
while for high activity of $l_r=180$, $S_0$ is non-monotonic. 
For $F_{D} < 0.0125$, which
corresponds to the clogged phase in which there is almost no motion,
$S_0$ is nearly zero.
Over the range $0.0125 \leq F_{D} < 0.2$,
the noise power is high and the system is in the
depinned or plastic regime with a coexistence of pinned
clusters and moving fluid.
There is a drop of $S_0$ to a low value for
$0.2 \leq F_{D} < 0.9$, corresponding
to the fluid phase.
Near $F_{D} = 1.0$, the noise power is high again in
the coexistence regime
of the fluid and cluster stripe phases,
while $S_0$ is low in the stripe crystal state for $F_{D} > 1.0$. 

From the behavior of the
noise and transport features,
we identify four dynamical phases in
the high activity system.
In phase I, 
the system is clogged with a large cluster and low mobility.
Phase II is the region in which
the fluctuating clogged fluid phase coexists with 
a fluid phase.
The pure fluid state is phase III, and phase II$_b$ is a
coexistence between the fluid phase and the stripe crystal phase.
Finally, at high drives, the stripe crystal phase IV emerges.
Large values of the noise power are 
associated with the coexistence phases in which the system
intermittently jumps between two distinct states.
The low activity system exhibits only
the 2D fluid phase III and phase IV$_{b}$, which is a 1D fluid.   
We can also examine the power spectra
directly, as plotted in Fig.~\ref{fig:6}(b) for 
$F_{D} = 0.03$ in the strongly fluctuating depinned clogged state 
and $F_{D} = 0.3$ in the fluid phase.
The dashed line is a fit to
a power law, $S(f) \propto f^{\alpha}$
with exponent $\alpha=-1.3$, for the strongly fluctuating state.
In non-active systems undergoing plastic depinning,
typical velocity fluctuation power spectra have a power law form
with an exponent in the range $-2.0 < \alpha < -1.0$ \cite{Reichhardt17}.
For the active system, we find that in the fluid state
the velocity noise is fairly white
and has an exponent $\alpha$ close to zero.
Additionally, the fluid phase
exhibits a characteristic peak at higher
frequencies that arises when the periodicity of the substrate
introduces a typical time between particle-obstacle collisions.
In phase II$_b$ we also find broad band or power law noise, while
the spectrum is white in phase IV.

The behavior of the high activity regime is
similar to what is found for flexible solids of interacting particles,
such as superconducting vortices or colloids,
driven over periodic arrays, and exhibits
similar peaks and dips in the
differential mobility as well as
peaks in the noise near the dynamical transitions 
\cite{Reichhardt17}.
In the active particle case,
the solid emerges at high activities due to the MIPS mechanism,
where the dense phase can be viewed as a solid with longer range correlations.  
In the low activity limit, the system
acts like a weakly correlated fluid driven over
a periodic substrate, where there are only short range correlations.
It is likely that as the run length is decreased, the behavior
would gradually change from solidlike to fluidlike. 

\section{Summary}
We have investigated the dynamics of active matter run-and-tumble disks 
moving through a periodic obstacle array under
varied external drives.
We consider the low activity or Brownian regime and
the high activity or motility induced phase separation regime where
clusters spontaneously form,
and we focus on systems
where the total density is well below the density at which
jamming would occur for passive disks.
For low activity, the velocity-force curves have a small nonlinear regime
at low drives where the system forms a 2D fluid, 
followed by a crossover to a linear regime at
higher drives where there is a 1D fluid flow state. The velocity noise
power is low for all drives. 
At high activity, the transport curves are strongly nonlinear and have
two peaks in the differential
mobility. The first peak corresponds to a depinning transition
from a clogged state to a flowing 2D fluid,
and the second peak is associated with the transition from a fluid to 
a density modulated crystalline stripe phase.
The moving stripe phase can display different patterns
depending on the lattice spacing and the overall density of the system. 
The velocity noise power also passes through peaks at the
transitions between the different flow states,
and the velocity noise has a broad band or $1/f$
character in the coexistence regime between two phases but is
white in the fluid phases.
We argue that the high activity system behaves much
like a solid depinning from a periodic substrate,
and that the solid-like behavior arises
due to the motility induced phase separation.
In contrast, the low activity state behaves like a weakly correlated fluid.  

\acknowledgments
We gratefully acknowledge the support of the U.S. Department of
Energy through the LANL/LDRD program for this work.
This work was supported by the US Department of Energy through
the Los Alamos National Laboratory.  Los Alamos National Laboratory is
operated by Triad National Security, LLC, for the National Nuclear Security
Administration of the U. S. Department of Energy (Contract No. 892333218NCA000001).

\bibliographystyle{eplbib}
\bibliography{mybib}


\end{document}